\documentclass[11pt]{article}
\usepackage{jheppub}
\usepackage{epsfig} 
\usepackage{amssymb}
\usepackage{amsmath}
\usepackage{epstopdf}
\usepackage{extarrows}

\usepackage{dsfont,bbm}
\usepackage{subfig}

\usepackage{shuffle}
\usepackage{booktabs}
\usepackage{array}

\graphicspath{{./figs/}}

\newcommand{\itbf}[1]{\textbf{\textit{#1}}}

\preprint{USTC-ICTS/PCFT-23-41}
\title{Color-Kinematics Duality with Minimal Deformation: Two-Loop Four-Gluon Amplitudes in Pure Yang-Mills Revisited}

\author[a,b]{Zeyu Li}
\emailAdd{lizeyu@itp.ac.cn}
\author[a,b,c,d,e]{and Gang Yang}
\emailAdd{yangg@itp.ac.cn}
\affiliation[a]{CAS Key Laboratory of Theoretical Physics, Institute of Theoretical Physics, \\Chinese Academy of Sciences, Beijing 100190, China}
\affiliation[b]{School of Physical Sciences, University of Chinese Academy of Sciences, Beijing 100049, China}
\affiliation[c]{School of Fundamental Physics and Mathematical Sciences, Hangzhou Institute for Advanced Study, UCAS, Hangzhou 310024, China}
\affiliation[d]{International Centre for Theoretical Physics Asia-Pacific, Beijing/Hangzhou, China}
\affiliation[e]{Peng Huanwu Center for Fundamental Theory, Hefei, Anhui 230026, China}

\abstract{
The conjectured duality between color and kinematics has significantly advanced our understanding of both gauge and gravitational theories. However, constructing numerators that manifest the color-kinematics (CK) duality, even for the two-loop four-gluon amplitude in pure Yang-Mills, has been challenging. In this paper, we revisit this amplitude and show that the difficulty of applying CK duality can be overcome by introducing a simple deformation. 
Our approach distinguishes itself from previous studies by maximizing the use of off-shell CK duality while maintaining a compact ansatz.
In particular, the deformation we introduce satisfies a subset of off-shell dual Jacobi relations.
The resulting numerators are presented in $d$-dimensionally Lorentz invariant local form and are applicable to all helicities of external gluons.
The solution we provide can be directly employed to construct the corresponding gravitational amplitude through double copy. 
Our findings suggest a novel and efficient strategy for constructing high-loop gauge and gravitational amplitudes using CK duality. 
}

\begin{document}

\maketitle

\setcounter{footnote}{0}

\section{Introduction}

Bern, Carrasco, and Johansson made a surprising discovery of a duality between color and kinematics \cite{Bern:2008qj, Bern:2010ue}, revealing a profound link between the kinematic structure and the color structure in gauge theories. This duality, which involves the full-color factors, holds the potential to transfer the advancements of the planar sector (\emph{i.e.}~in the large $N_c$ limit) to the full-color sector. Furthermore, it enables the construction of gravity amplitudes directly from Yang-Mills amplitudes, provided that the latter are organized in a way that respects the duality. This is commonly referred to as the double copy property, a concept that generalizes the KLT relation \cite{Kawai:1985xq}. 

While the color-kinematics (CK) duality at the tree level has been established using both string theory and gauge theory methods \cite{BjerrumBohr:2009rd, Stieberger:2009hq, Feng:2010my, Bjerrum-Bohr:2010pnr, Mafra:2011kj}, the duality at the loop level remains conjectural and has only been substantiated through specific examples. 
Explicit CK-dual solutions at loop level have been found in a wide class of theories for both amplitudes and form factors \cite{Bern:2010tq, Carrasco:2011mn,Bern:2012uf, Du:2012mt, Oxburgh:2012zr, Yuan:2012rg,Boels:2012ew, Boels:2013bi, Bjerrum-Bohr:2013iza, Bern:2013yya, Ochirov:2013xba, Mafra:2015mja,
He:2015wgf, Mogull:2015adi, Yang:2016ear, Chiodaroli:2017ngp,Boels:2017skl, He:2017spx,Johansson:2017bfl,Jurado:2017xut, Geyer:2017ela, Faller:2018vdz,
Lin:2020dyj,Edison:2020uzf,Lin:2021qol,Lin:2021lqo, Li:2022tir}.
See \cite{Bern:2019prr,Bern:2022wqg} for an extensive review of the color-kinematics duality and double copy.

Despite significant efforts, it remains a great challenge to understand to which extent the CK-duality holds. 
State-of-the-art constructions that manifest CK duality at high-loop levels include the four-loop four-point amplitude \cite{Bern:2012uf} and the five-loop Sudakov form factor \cite{Yang:2016ear} in ${\cal N}=4$ SYM. 
In non-supersymmetric gauge theories, constructing CK-dual loop integrands has proven particularly difficult, where the best understood high-loop solutions are the two-loop four-gluon and five-gluon amplitudes with all helicities equal \cite{Bern:2013yya, Mogull:2015adi}. 

In solving the difficulty of constructing global CK-dual solutions, a common approach is to enlarge the ansatz beyond the simple local form.
Notably, to achieve the CK-duality in the all-plus-helicity five-gluon two-loop amplitude in pure YM, numerators containing twelve powers of loop momenta are required, which significantly exceeds the complexity of standard Feynman diagrams \cite{Mogull:2015adi}. 
In the case that enlarging the ansatz is too complicated to work with,
one may opt to relax the requirement of global CK duality to only necessitate its validity at the level of cut integrand, as demonstrated in the two-loop four-gluon amplitude in $d$ dimensional kinematics in pure YM \cite{Bern:2015ooa}. 
However, these approaches result in a substantial increase in the number of parameters.

In this paper, we propose a novel strategy to apply CK duality by revisiting the two-loop four-gluon amplitude in pure YM theory. 
Our primary objective is to identify and understand the specific obstructions to constructing a global CK-dual solution.
With this understanding, we explore whether it is possible to implement a minor amendment—through minimal deformation—to efficiently construct the loop integrand while still utilizing CK-duality relations. As demonstrated by the four-gluon amplitude example, the answer is affirmatively positive.

Concretely, we begin with the most basic ansatz in a fully $d$-dimensional Lorentz-invariant local form.
Such an ansatz, manifesting graph symmetries and natural power-counting constraints,
cannot satisfy both global CK duality and unitarity cuts, an issue previously noted in \cite{Bern:2015ooa}. 
The primary obstruction is that CK duality conflicts with the ladder-type two-double cut, although it is compatible with other cuts. 
Rather than abandoning the complete set of dual Jacobi relations, we navigate around this obstruction by introducing a minimal deformation that corrects the two-double cut without affecting other cuts. 

Importantly, we also propose that the deformation satisfies a subset of off-shell dual Jacobi relations. These relations are confined to topologies that contribute to the ladder-type two-double cut. With these relations, one can choose one single numerator—the planar double-box topology—as the master numerator for the deformation.
The ansatz of the deformation can thus be given in a very compact form.
Surprisingly, we find in the final physical solutions, the deformation can be given in a remarkably simple form, see Eq.\,\eqref{eq:deformed_1}. 
This suggests that the global off-shell CK duality is only slightly violated.
We also discuss the structure of the solution space for the deformation in detail.

This paper is organized as follows. 
In Section~\ref{sec:reviewCK}, we briefly review CK duality and the construction procedure, including the unitarity method.
In Section~\ref{sec:amp2loop}, we examine the global CK duality for the two-loop four-gluon amplitudes in pure YM and discuss the difficulties in finding physical solutions.
Section~\ref{sec:deformation} presents the main result of this paper, where we introduce the deformation and explore the solution space.
A summary and discussion of our findings are given in Section~\ref{sec:discussion}.

The explicit solutions are also included in the ancillary files.

\section{Review}
\label{sec:reviewCK}
In this section, we review the basic concept of CK duality and the general strategy for constructing CK-dual integrands at the loop level.

\subsection{CK duality}

The CK duality conjectures that there exists a cubic graph
representation of amplitudes in which the kinematic numerators satisfy the same equations of Jacobi relations for the color factors.

\begin{figure}[t]
	\centerline{\includegraphics[height=2.7cm]{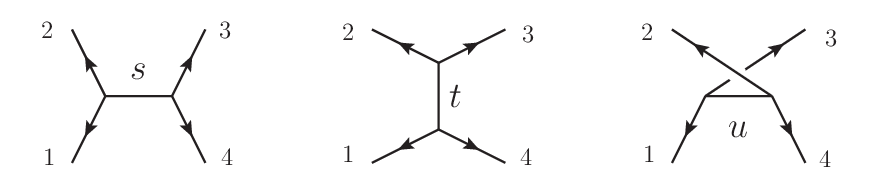} } 
	\caption{Trivalent graphs for the four-point tree amplitude.} 
	\label{treeA4}
\end{figure}

The basic and key example is the four-gluon tree amplitude which can be represented as
\begin{equation}\label{eq:fullcolortreeA4}
	\itbf{A}_4^{\rm tree}=g^2\left(\frac{c_sn_s}{s}+\frac{c_tn_t}{t}+\frac{c_un_u}{u}\right)\,,
\end{equation}
which is expanded in terms of three trivalent topologies in Figure~\ref{treeA4}. The color factor is defined as
\begin{equation}\label{color_def}
	c_s=\tilde{f}^{a_1a_2s}\tilde{f}^{sa_3a_4},\; \, c_t=\tilde{f}^{a_2a_3t}\tilde{f}^{ta_4a_1}, \;\, c_u=\tilde{f}^{a_1a_3u}\tilde{f}^{ua_2a_4},
\end{equation}
where $\tilde{f}^{abc}$ is structure constant:
\begin{equation}\label{def:StrucCons}
	\tilde{f}^{abc}\text{=i}\sqrt{2}f^{abc}\text{=tr([}T^a,T^b\text{],}T^c\text{)}\,,
\end{equation}
in which $T^a$ are group generators and are normalized by $\text{tr(}T^aT^b\text{)=}\delta ^{ab}$. In this paper, we will consider pure Yang-Mills theory and the gauge group will be specified as $SU(N_c)$. We will also abbreviate $\tilde{f}$ as $f$ in the rest of the paper.

Clearly, the color factors $c_{s,t,u}$ satisfy the Jacobi identity
\begin{equation}\label{stu_Jacobi}
	c_s= c_t + c_u \,,
\end{equation}
which is straightforward from the Jacobi relation of structure constants. However, The CK duality requires the numerators, which contain all the physical information, also satisfy the same relation as the color factors:
\begin{equation}
	n_s= n_t + n_u \,.
\end{equation}
This relation be referred to as ``dual Jacobi relation'' or ``CK-dual relation''. 
In this four-point tree amplitude case, it is easy to obtain $n_{s,t,u}$ using Feynman rules and check that they indeed satisfy the CK relation. For high-point tree amplitudes, it is also possible to construct numerators satisfying the CK duality. At loop level, however, the duality remains a conjecture and we need to check it case by case.

The general form of a $L$-loop $m$-point amplitude can be written as:
\begin{equation}
	\itbf{A}_m^{(L)}=i^L g^{m-2+2L} \sum\limits_{\sigma_m} \sum\limits_{\Gamma_i}  \int \prod\limits_{j=1}^{L} \frac{d^{D}l_j}{(2\pi)^D} \frac{1}{S_{i}} \frac{C_i N_i}{\prod_a  D_{i,a}}.
\end{equation}
The first summation of $\sigma_m$ runs over the $m!$ permutations of the external legs. The summation over $\Gamma_i$ means to sum over all
possible trivalent graphs and the $S_i$ will remove all the overcounting that comes from the summation of external legs. $C_i$ and $N_i$ corresponds to the color factor and kinematic numerator of the $i$th trivalent graph, and $1/D_{i,a}$ denotes as the $a$th propagator of the $i$th graph. 

\begin{figure}[t]
	\centerline{\includegraphics[width=0.99\linewidth]{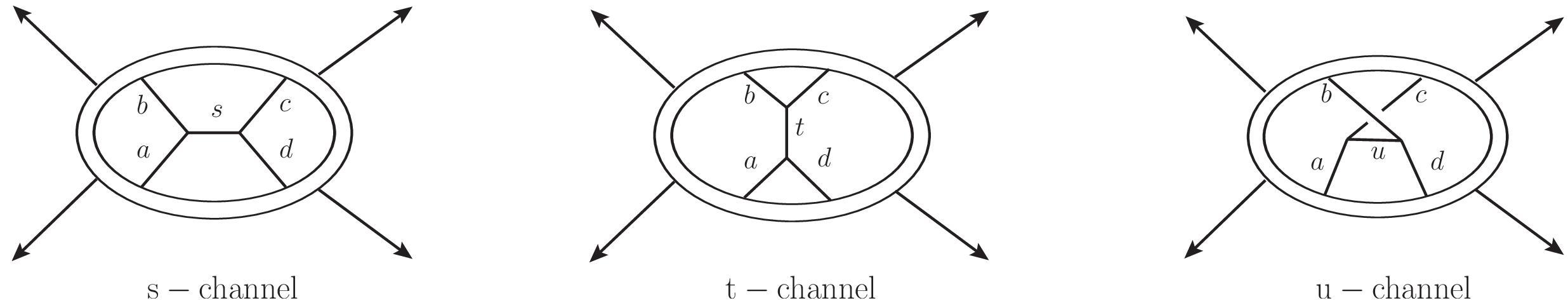} } 
	\caption{Trivalent graphs and CK-relations at loop level.} 
	\label{loop_ck_relation}
\end{figure}

To generalize the CK duality into loop level, we can embed the relation \eqref{color_def} and \eqref{stu_Jacobi} into a loop trivalent graph, see Figure~\ref{loop_ck_relation}. Note that except for the four-point sub-digram, the rest part of the three diagrams are the same, so their color factors take the form:
\begin{equation}
	C_s= f^{abs}f^{scd}\prod{f}, \quad C_t= f^{bct}f^{tda}\prod{f}, \quad C_u= f^{acu}f^{ubd}\prod{f},
\end{equation}
apparently, they satisfy the Jacobi identity
\begin{equation}\label{loop_stu_Jacobi}
	C_s= C_t + C_u,
\end{equation}
and CK duality requires their corresponding numerators to satisfy:
\begin{equation}\label{loop_stu_numer_Jacobi}
	N_s(\{l_a,l_b,l_s\},\{-l_s,l_c,l_d\})= N_t(\{l_d,l_a,l_t\},\{-l_t,l_b,l_c\}) + N_u(\{l_c,l_a,l_u\},\{-l_u,l_d,l_b\}),
\end{equation}
which is the dual Jacobi relation at the loop level. For each propagator in each trivalent graph, we can write down such a CK-dual identity. 

An important goal is to construct the $N_i$ which satisfies all the CK identities. This has the advantage that once the CK-dual representation is found, the gauge amplitude can be used to construct a corresponding gravitational amplitude via double copy. Moreover, as we will see, imposing the duality constraints makes it possible to significantly simplify the loop construction.

In practical, we can construct the CK ansatz for $N_i$ by following standard steps \cite{Bern:2012uf, Boels:2012ew, Carrasco:2015iwa}:
\begin{enumerate}
\item 
Generate all the trivalent diagrams.  
\item 
Generate all CK relations and find ``master topologies''. The ``master topologies'' represent a minimal set of topologies which can deduce all other diagrams through dual Jacobi relations. The choice of master topologies is generally not unique.
\item 
Construct an ansatz for the numerators of master topologies. Once we have the ansatz for master topologies, we can obtain all other numerators by CK-dual relations. 
\item 
Apply the symmetry constraints and require each numerator to reflect the same symmetry property as its topology. 
\end{enumerate}
Once we obtain the CK ansatz for integrand, the next task is to apply unitarity cuts to the integrand to ensure it provides a physical result, which will be reviewed in the next subsection.

\subsection{Full-color $d$-dimensional unitarity-cut method}\label{sec:unitarity}
In this subsection, we review the unitarity method for loop integrand construction. Since we are studying the pure YM theory in this paper, we stress that one needs to apply $d$-dimensional full-color cuts to get the complete result.

The central idea of the unitarity method \cite{Bern:1994zx,Bern:1994cg,Britto:2004nc} is that by setting internal propagators to be on-shell as
\begin{equation}\label{eq:oneloopNuExpansion}
	\frac{i}{l^2} \stackrel{\rm cut}{\longrightarrow} 2\pi\delta_{+}(l^2) \,,
\end{equation}
the amplitude will be factorized as products of lower-order amplitudes, such as
\begin{equation}\label{general_cut_equation}
	\itbf{A}^{(l)} |_{\rm cuts} =\int {\rm dPS} \sum_{\rm physical \; states} \prod_{i} \itbf{A}^{(0)}_{n_i} \,,
\end{equation}
where $\itbf{A}^{(0)}_{n_i}$ is full-color tree amplitude with $n_i$ external legs. The physical result must be consistent in all possible cut channels.

Since we consider pure YM theory, it is necessary to consider $d$-dimensional cuts. The tree amplitudes are calculated by Feynman diagrams and all the expressions appear as Lorentz products which are valid in $d$ dimension. 
To perform the sum of the helicities for all possible internal cut states we use
\begin{equation}\label{helicity_sum}
	\sum_{\rm{physical \; states}} \varepsilon^{\mu}(l) \varepsilon ^{\nu} (l) = \eta^{\mu \nu} - \frac{l^{\mu} \xi^{\nu} + l^{\nu} \xi^{\mu}}{l \cdot \xi} \,,
\end{equation}
where the $\xi^{\mu}$ is a light-like reference momentum and the result after summation should be
independent of it.
The amplitude thus computed is in the conventional dimensional regularization scheme where all gluons (whether internal or external) are taken to be $d$ dimensional, and the result applies to all possible helicity configurations of external gluons in the four-dimensional limit.

In practice, it will be convenient to perform color decomposition, which will allow us to use color-stripped tree amplitude as building blocks.
On the RHS of \eqref{general_cut_equation}, we decompose each full-color tree amplitude into color-stripped amplitude as
\begin{equation}\label{full_color_cut_tree_product}
	\prod_{i} \itbf{A}^{(0)}_{n_i} = \prod_{i}\sum_{\sigma_{n_i}}{\rm tr}(\sigma_{n_i})A^{(0)}(\sigma_{n_i})= \sum_{\sigma_1 \sigma_2 ... \sigma_{n_i}} {\rm tr}(\sigma_1){\rm tr}(\sigma_2)...{\rm tr}(\sigma_{n_i}) A^{(0)}(\sigma_1)A^{(0)}(\sigma_2)...A^{(0)}(\sigma_{n_i}).
\end{equation}
On the LHS, we decompose the color factor of each topology into the same product of multi-trace bases :
\begin{equation}
	\sum_{\Gamma_j} \frac{C_j N_j}{\prod_a  D_{j,a}}\Bigg|_{cut} = \sum_{\Gamma_j} \sum_{\sigma_1 \sigma_2 ... \sigma_{n_i}} {\rm tr}(\sigma_1){\rm tr}(\sigma_2)...{\rm tr}(\sigma_{n_i}) \frac{N_j}{\prod_a  D_{j,a}}\Bigg|_{cut}.
\end{equation}
After decomposing both sides of \eqref{general_cut_equation} into the same multi-trace bases, we can compare the coefficient of each base and apply the unitarity constraint.

\begin{figure}[t]
	\centerline{\includegraphics[height=3.5cm]{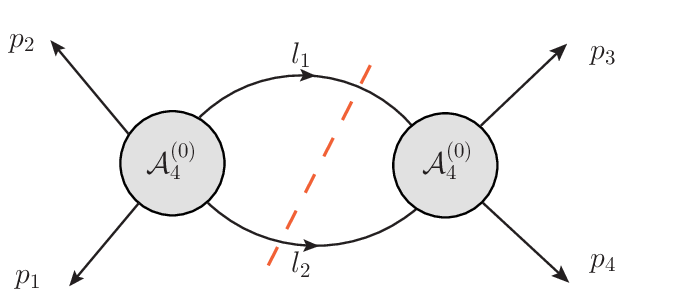} } 
	\caption{Double cut for the one-loop four-point amplitude.} 
	\label{fig:1-loop_cut}
\end{figure}

\begin{figure}[t]
	\centerline{\includegraphics[width=0.99\linewidth]{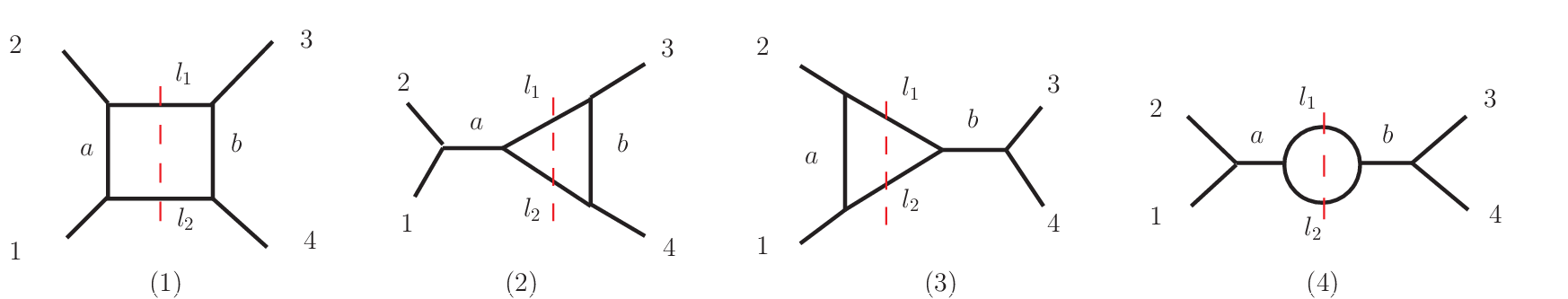} } 
	\caption{Diagrams contributing to the double cut in Figure~\ref{fig:1-loop_cut}.} 
	\label{fig:1_loop_cut_topology}
\end{figure}

As a concrete simple example, we consider the double cut for one-loop four-point amplitude in Figure~\ref{fig:1-loop_cut}. The general equation of \eqref{general_cut_equation} now becomes
\begin{equation}
	\itbf{A}_4^{(1)}(1,2,3,4) \Big| _{\rm s-cut} = \int {\rm dPS} \sum_{\rm physical \; states}  \itbf{A}^{(0)}_{4}(1,2,l_1,l_2) \itbf{A}^{(0)}_{4}(3,4,-l_2,-l_1)  \,.
\end{equation}
On the RHS, we decompose each $\itbf{A}^{(0)}_{4}$ into color-stripped amplitude and extract the coefficient of an explicit base ${\rm tr}(1,2,l_1,l_2){\rm tr}(3,4,l_2,l_1)$
\begin{equation} 
	\sum_{\rm physical \; states}  A^{(0)}_{4}(1,2,l_1,l_2) A^{(0)}_{4}(3,4,-l_2,-l_1)  \,.
\end{equation}
On the LHS, we first select out all the trivalent diagrams that contribute to this cut in Figure~\ref{fig:1_loop_cut_topology} and extract the contribution to ${\rm tr}(1,2,l_1,l_2){\rm tr}(3,4,l_2,l_1)$ for each color factor
\begin{equation}
	\begin{aligned}
		C_1 &= f^{1al_2}f^{2l_1a}f^{l_1 3 b}f^{4 l_2 b} \longrightarrow (+1){\rm tr}(1,2,l_1,l_2){\rm tr}(3,4,l_2,l_1),\\
		C_2 &= f^{12a}f^{al_1l_2}f^{l_1 3 b}f^{4 l_2 b} \longrightarrow (+1){\rm tr}(1,2,l_1,l_2){\rm tr}(3,4,l_2,l_1),\\
		C_3 &= f^{1al_2}f^{2l_1a}f^{l_2 b l_1}f^{b34} \longrightarrow (-1){\rm tr}(1,2,l_1,l_2){\rm tr}(3,4,l_2,l_1),\\
		C_4 &= f^{12a}f^{al_1l_2}f^{l_1 b l_2}f^{b34} \longrightarrow (+1){\rm tr}(1,2,l_1,l_2){\rm tr}(3,4,l_2,l_1) \,. \\
	\end{aligned}
\end{equation}
Combine LHS and RHS we have:
\begin{align}
&	\begin{gathered}
		\includegraphics[height=1.8cm]{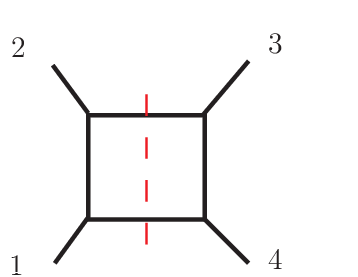}
	\end{gathered} \hspace{-6mm}  \times \;  N_1 \; + 
\begin{gathered}
\includegraphics[height=1.8cm]{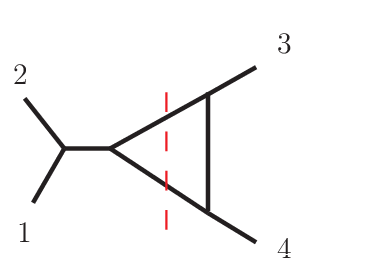}
\end{gathered} \hspace{-5.8mm} \times \;  N_2 \; + \hspace{-2.5mm}
\begin{gathered}
\includegraphics[height=1.8cm]{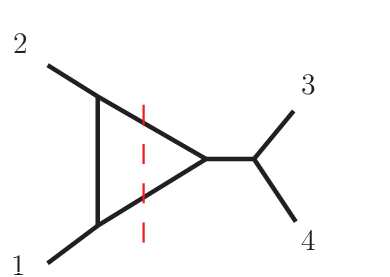}
\end{gathered} \hspace{-4.7mm} \times \; (-1) N_3 \; +
\begin{gathered}
\includegraphics[height=1.3cm]{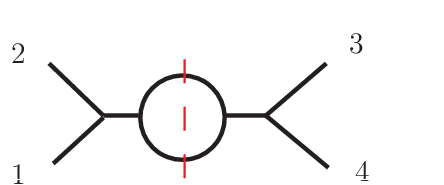}
\end{gathered} \hspace{-5.5mm} \times \;  N_4 \nonumber\\
&=\sum_{\rm physical \; states}  A^{(0)}_{4}(1,2,l_1,l_2) A^{(0)}_{4}(3,4,-l_2,-l_1)  \,.
\end{align}
After performing the helicity sum \eqref{helicity_sum}, both sides of the above equation are rational functions involving the Lorentz product of $\{ l_i, p_i, \varepsilon_i\}$, as well as the dimension parameter $d$.
It is straightforward to consider other ordering of trace bases, including non-planar cuts.

\section{Difficulty of solving an ansatz with global CK duality}
\label{sec:amp2loop}

In this section, we explore the integrand construction for two-loop four-point pure YM amplitude in $d$ dimensions using global CK duality. As we will see, the global CK-dual integrand in local ansatz form is not compatible with all the unitarity cuts. We will discuss the difficulty and the possible solutions.

\subsection{Ansatz with global CK duality}

In Section~\ref{sec:reviewCK} we briefly review the general strategy for constructing the CK integrand. Here we concretize each step for the two-loop four-point amplitude.

\begin{figure}[t]
	\centerline{\includegraphics[width=1.0\linewidth]{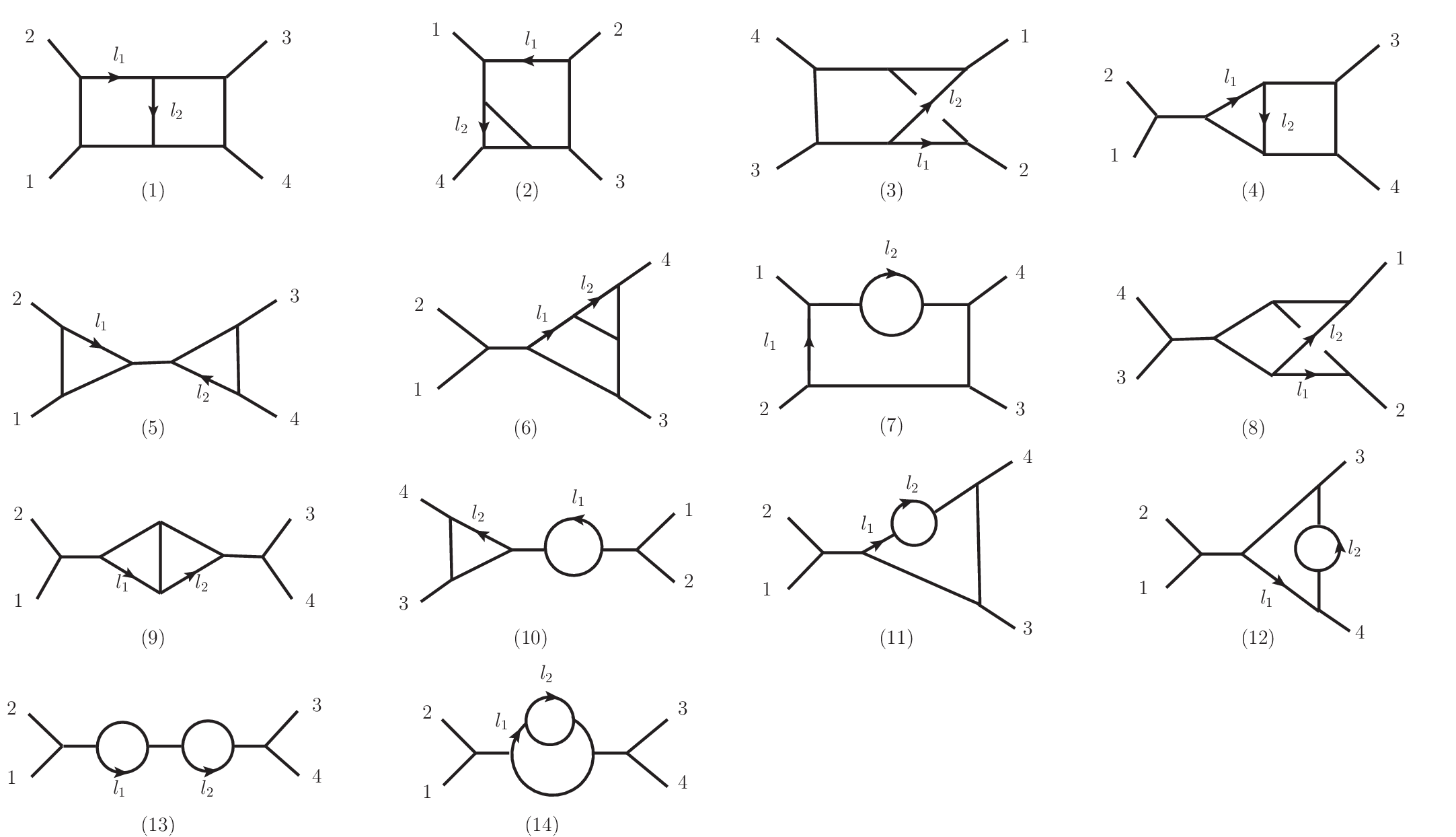} } 
	\caption{All trivalent diagrams for the two-loop four-point amplitude. } 
	\label{fig:2_loop_all_topology}
\end{figure}

\paragraph{1) Generating trivalent diagrams.} As the first step, we generate all the trivalent diagrams for two-loop four-point amplitude. All these diagrams are collected in Figure~\ref{fig:2_loop_all_topology}.
Topologies with tadpoles or massless bubbles are excluded since they contain scaleless integral and vanish after integration.\footnote{More precisely, we will exclude the CK relations that involve these topologies, so they will not affect the CK integrand at all.}

\paragraph{2) Generating CK relations.} After obtaining all trivalent diagrams, we can do the Jacobi operation on each propagator in each graph and generate all the CK relations. Solving the CK relations we find all other numerators can be deduced by diagrams (1) and (2) in Figure~\ref{fig:2_loop_all_topology}, so we choose them to be ``master topologies". Their numerators are denoted as $n_1$ and $n_2$. We list the CK relations that can deduce the rest numerators below:
\begin{equation}\label{eq:dualjacobirelationforni}
	\begin{aligned}
		n_3&=n_1[p_1,p_2,p_3,p_4,-l_2-p_2,l_1-p_2]+n_2[p_4,p_3,p_2,p_1,-l_1-l_2-p_3,-l_2+p_1]\\
		n_4&=n_1-n_1[p_3,p_4,p_2,p_1,l_1-l_2+p_1+p_2,-l_2]\\
		n_5&=-n_1[p_1,p_2,p_3,p_4,l_1,l_1-l_2+p_1+p_2]+n_1[p_1,p_2,p_4,p_3,l_1,l_1+l_2]\\
		n_6&=n_2[p_1,p_2,p_3,p_4,l_1+p_1,l_2]-n_2[p_3,p_1,p_2,p_4,-l_1-p_1-p_2,-l_2-p_1-p_2-p_3]\\
		n_7&=-n_2[p_1,p_2,p_3,p_4,l_1,l_2]-n_2[p_1,p_2,p_3,p_4,l_1,l_1-l_2-p_1]\\
		n_8&=n_3-n_3[p_1,p_2,p_4,p_3,l_1,l_2]\\
		n_9&=-n_4[p_1,p_2,p_3,p_4,l_1,l_1-l_2]+n_4[p_1,p_2,p_4,p_3,l_1,l_1-l_2]\\
		n_{10}&=-n_4[p_1,p_2,p_3,p_4,l_1,l_1+l_2+p_1+p_2]-n_4[p_1,p_2,p_3,p_4,-l_1-p_1-p_2,-l_1+l_2]\\
		n_{11}&=-n_4[p_1,p_2,p_3,p_4,-l_2-p_1-p_2,l_1-l_2]-n_4[p_1,p_2,p_3,p_4,-l_1+l_2-p_1-p_2,l_2]\\
		n_{12}&=-n_6[p_1,p_2,p_3,p_4,l_1,l_1-l_2]-n_6[p_1,p_2,p_3,p_4,l_1,l_2-p_1-p_2-p_3]\\
		n_{13}&=n_9+n_9[p_1,p_2,p_3,p_4,-l_1-p_1-p_2,l_2]\\
		n_{14}&=n_9[p_1,p_2,p_3,p_4,l_1-l_2,l_1]+n_9[p_1,p_2,p_3,p_4,-l_2-p_1-p_2,-l_1-p_1-p_2] \,,
	\end{aligned}
\end{equation}
where $n_i$'s are abbreviation of $n_{i}[p_1,p_2,p_3,p_4,l_1,l_2]$. These CK-dual relations are satisfied throughout all the topologies in Figure~\ref{fig:2_loop_all_topology}, and we will refer to them as ``global CK-dual relations".

\paragraph{3) Constructing numerator ansatz.} Now we construct an ansatz for the two master numerators. The ansatz is linear combination of monomials $M_{k}$:
\begin{equation}
	\label{master_ansatz}
	n_m =  \sum_{k} a_{mk} M_{k} \,, \qquad m = 1,2 \,,
\end{equation}
where the monomials $M_{k}$ are built by product of following basis:
\begin{equation}\label{lorntzproduct_basis}
	\{ \varepsilon_i \cdot \varepsilon_j, \; \varepsilon_i \cdot p_j, \; \varepsilon_i \cdot l_\alpha, \; p_i \cdot l_\alpha, \; l_\alpha \cdot l_\beta \;, p_1 \cdot p_2 \;, p_2 \cdot p_3 \} \;
\end{equation}
with $i,j = 1,2,3,4$ and $\alpha,\beta = 1,2$. 
We choose a set of bases for $M_k$ by eliminating non-independent ones under
momentum-conservation, on-shell, and transversality conditions:
\begin{equation}
	\label{momentum_conservation_equation}
	p_4=-p_1-p_2-p_3, \;\; p_i^2 = 0, \;\; \varepsilon_{i} \cdot p_i =0.
\end{equation}
Besides, monomial $M_{k}$ should obey the following features: 
(1) Each $M_{k}$ have mass dimension six.
(2) $M_{k}$ should depend on each polarization vector $\varepsilon_{i}$ linearly.
Since they are polynomials of Lorentz products, they are also free with poles. 
One such example is $(\varepsilon_1 \cdot \varepsilon_2)(\varepsilon_3 \cdot \varepsilon_4)(p_1 \cdot p_2)(p_1 \cdot l_1)(p_1 \cdot l_2)$. 
With these features satisfied, the two master numerators introduce 20020 terms in total.\footnote{Note that in \cite{Bern:2015ooa}, the ansatz for master numerators are a bit smaller since they made restrictions to the power of loop momentum. Here we do not impose such restrictions. Such a choice of ansatz was also discuessed in \cite{Edison:2023ulf}.}
Here we also mention that the coefficients $a_{mk}$ will depend on the dimension parameter $d$. All other numerators are obtained using the set of dual Jacobi relations \eqref{eq:dualjacobirelationforni}.

\paragraph{4) Applying symmetry constraints.} We demand each numerator reflects the symmetries of its topology:
\begin{equation}\label{symmetry}
	\hat{S}[C_i n_i] = C_i n_i,
\end{equation}
where $\hat{S}$ denotes symmetry operator, which will act on both $C_i$ and $n_i$. 
For instance, the double box diagram possesses two symmetries: vertical flip and horizontal flip. It is easy to see that the color factor remains unchanged and the numerator is required to satisfy
\begin{equation}\label{eqs: double_box_symmetry}
	n_1=n_1[p_2,p_1,p_4,p_3,-p_1-p_2-l_1,-l_2] 
		=n_1[p_4,p_3,p_2,p_1,l_2-l_1,l_2] \,,
\end{equation}
and similar for other topologies. 
After applying the constraints of symmetry and global CK-dual relations, we find that the number of parameters reduce to 1382.

Up to now, we have constructed the global CK relations satisfied integrand for two-loop four-point amplitude and we will apply the unitarity constraints in the next subsection.

\subsection{Problem with unitarity constraints}

Given the ansatz, we now apply unitarity cuts. 

\begin{figure}[t]
	\centerline{\includegraphics[height=2.7cm]{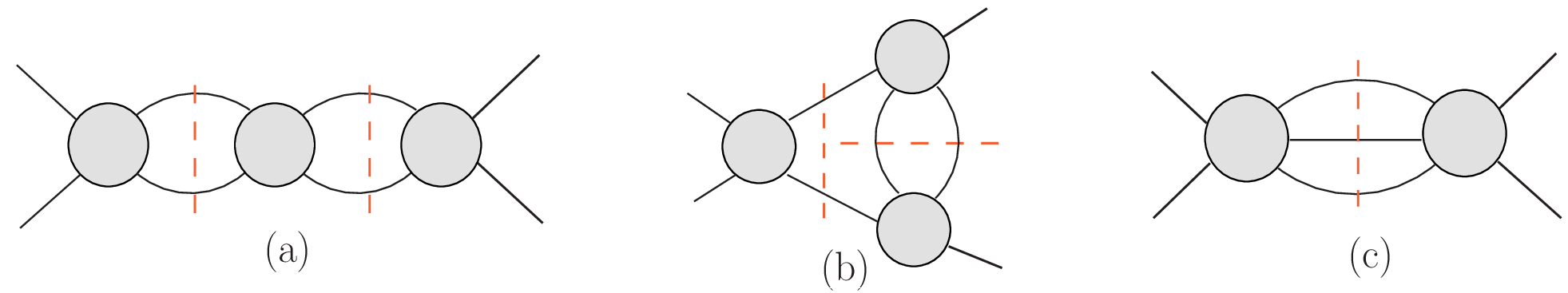} } 
	\caption{A spanning set of cuts for two-loop four-point amplitude.} 
	\label{fig:2_loop_spanning_cuts}
\end{figure}

For the two-loop four-point amplitude, a spanning set of unitarity cuts is displayed in Figure~\ref{fig:2_loop_spanning_cuts}.\footnote{The spanning set of cuts contain both planar and non-planar cuts. 
In the following discussion, the check of the cuts can be first taken as for the planar cuts. Due to the CK-duality relations, the solutions will automatically satisfy non-planar cuts in the end, see \emph{e.g.}~\cite{Bern:2008qj, Badger:2015lda}. We will mention this check at the end of Section~\ref{sec:deformsolution}. }
We find that cut-(b) and cut-(c) in Figure~\ref{fig:2_loop_spanning_cuts} are compatible with CK-dual relations, while cut-(a) is not. 
The same observation was pointed out in \cite{Bern:2015ooa,Edison:2023ulf}.
This shows that the global CK-dual integrand with local ansatz form satisfying all symmetry constraints can not deduce physical result.

One may try to solve this problem by following ideas.

\paragraph{1)} One can simplify the problem by considering helicity amplitudes. In a specific helicity amplitude, the physical result may be much simpler so it will be easier to realize CK duality. For two-loop four-point amplitude, helicity configuration can be $(++++)$, $(+++-)$, $(++--)$ and $(+-+-)$. Case of $(++++)$ has already been constructed in \cite{Bern:2013yya}.

\paragraph{2)} One can try to enlarge ansatz such as introducing the non-local property to numerators and increasing the power of loop momenta. This method has been used in \cite{Mogull:2015adi} to realize the CK duality for pure Yang-Mills two-loop five-point amplitude with identical helicities.

\begin{figure}[t]
	\centerline{\includegraphics[height=2.5cm]{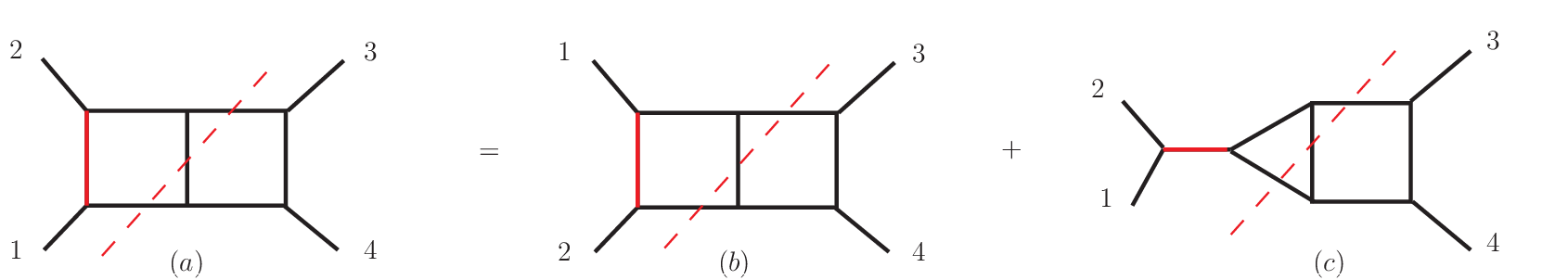} } 
	\caption{An example for CK relation in unitarity cut.} 
	\label{fig:cut_CK_example}
\end{figure}

\paragraph{3)} One can relax the constraints of CK-dual identities as in \cite{Bern:2015ooa} by demanding that they hold only on a spanning set of cuts without losing the double-copy property. For example, the three numerators in Figure~\ref{fig:cut_CK_example} should satisfy:
\begin{equation}
	\label{cut_CK_relation_equation}
	(n_{a}-n_{b}-n_{c})\big|_{\rm cut}=0 \,.
\end{equation}
However, in such a strategy, one would give up using the global CK-dual identities to obtain the whole integrand. Instead, an ansatz for each diagram in Figure~\ref{fig:2_loop_all_topology} has to be made, and the complete ansatz is much larger than the usual strategy utilizing master topologies. For example, the ansatz in \cite{Bern:2015ooa} contains 120904 parameters for all 14 topologies in total. After imposing symmetry constraints, 28204 parameters remain. While taking unitarity cuts, the cut CK-dual identities like \eqref{cut_CK_relation_equation} are imposed.
It is not hard to see that the method would be difficult for more complicated cases. With the growth of loops and external momentums, the scale of ansatz will increase rapidly since all the topologies need to be treated separately.\footnote{For example, a similar strategy for the three-loop four-gluon amplitude would require an ansatz that involves more than millions of parameters, and it would be important to reduce the number of topologies for ansatz construction.}

In this paper we propose a new strategy: we start with a global CK-dual integrand and then introduce simple ``deformations" which can also be related by CK-dual relations.  
In the next section, we will provide a detailed application of this strategy.

\section{Physical solution with deformation}\label{sec:deformation}

This section presents the major results of the paper. We first discuss the strategy of introducing deformation for the CK-dual numerators. Then we provide the explicit solution. Finally, we discuss the solution space for the deformations.

\subsection{General idea of deformation}\label{sec:generalidea}

\begin{figure}[t]
	\centerline{\includegraphics[width=0.99\linewidth]{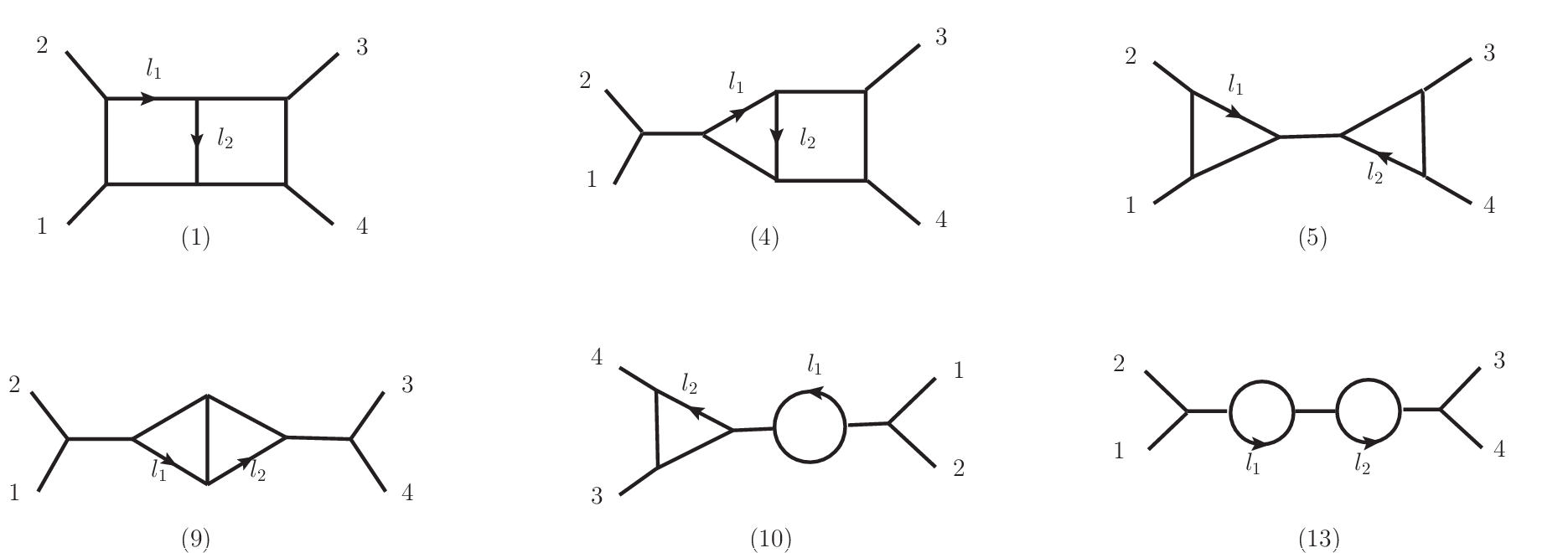} } 
	\caption{Topologies that will contribute to cut-(a) in Figure~\ref{fig:2_loop_spanning_cuts}  thus need to be deformed.} 
	\label{fig:2_loop_deformed_topology}
\end{figure}

We would like to work based on $n_i$ and construct a physical solution that passes all unitarity cuts. 
Let us first recall the property of $n_i$ that we have obtained in the previous section based on CK-dual construction. 
\begin{itemize}
\item $n_i$ have the local ansatz form and satisfy the symmetry properties.
\item $n_i$ satisfy the global set of CK-dual relations \eqref{eq:dualjacobirelationforni}.
\item The integrand of $n_i$ satisfies cut-(b) and cut-(c) in Figure~\ref{fig:2_loop_spanning_cuts}.
\item The solution space satisfying the above properties for $n_i$ contains 398 parameters.
\end{itemize}
Since the main problem comes from the cut-(a) in Figure~\ref{fig:2_loop_spanning_cuts}, we will concentrate on the topologies that can affect cut-(a), which are collected in Figure~\ref{fig:2_loop_deformed_topology}.

The main idea is to introduce certain deformation $\Delta_{i}$ to the numerators of these topologies, such that the deformation $\Delta_{i}$ plus $n_{i}$ together will satisfy the unitarity-cut constraints.
Numerators whose topology does not appear in Figure~\ref{fig:2_loop_deformed_topology} remain unchanged, or equivalently, their deformations are set to be zero. 
Concretely, we define the full set of physical numerators as
\begin{equation}\label{eq:deformed_numerators}
	N_i = \left\{ \begin{matrix} & n_i +\Delta_i, & \qquad i=1,4,5,9,10,13, \\ & n_i , & \textrm{others.} \end{matrix} \right.
\end{equation} 
The deformation $\Delta_i$ should vanish in cut-(b) and cut-(c) since the $n_i$ part already satisfies these cuts.

Clearly, the set of numerators in \eqref{eq:deformed_numerators} will break the global CK duality since we only make non-zero deformations to a subset of topologies. On the other hand, we would like the deformed numerators to have the important property that: they can be used for the double-copy construction. 
To achieve this, one should at least require $N_{i}$ to satisfy CK relations under all unitarity cuts, as in \cite{Bern:2015ooa}. 
Since all $n_i$ are constructed by satisfying (global) CK relations, the deformations $\Delta_i$ must also satisfy dual Jacobi identities under cuts.

Rather than just imposing CK relations with cuts, we propose that $\Delta_{i}$ should satisfy a sub-set of \emph{off-shell} dual Jacobi relations. 
This is one key point of our proposal, and as we see below, it will significantly simplify the calculation. 

We consider dual Jacobi relations that involve only the topologies in Figure~\ref{fig:2_loop_deformed_topology}. 
It turns out that one can choose the double box diagram as the master topology for the deformation. Given its numerator deformation $\Delta_1$, the deformation for other five topologies in Figure~\ref{fig:2_loop_deformed_topology} can be determined through following dual Jacobi relations:
\begin{align}\label{eq:2_loop_deformations_CK}
		\Delta_4&=\Delta_1-\Delta_1[p_3,p_4,p_2,p_1,l_1-l_2+p_1+p_2,-l_2] \nonumber\\
		\Delta_5&=-\Delta_1[p_1,p_2,p_3,p_4,l_1,l_1-l_2+p_1+p_2]+\Delta_1[p_1,p_2,p_4,p_3,l_1,l_1+l_2] \\
		\Delta_9&=-\Delta_4[p_1,p_2,p_3,p_4,l_1,l_1-l_2]+\Delta_4[p_1,p_2,p_4,p_3,l_1,l_1-l_2] \nonumber\\
		\Delta_{10}&=-\Delta_4[p_1,p_2,p_3,p_4,l_1,l_1+l_2+p_1+p_2]-\Delta_4[p_1,p_2,p_3,p_4,-l_1-p_1-p_2,-l_1+l_2] \nonumber\\
		\Delta_{13}&=\Delta_9+\Delta_9[p_1,p_2,p_3,p_4,-l_1-p_1-p_2,l_2] \,. \nonumber
\end{align}
Note that we do not impose any cut conditions for these relations.

\begin{figure}[t]
	\centerline{\includegraphics[height=2.6cm]{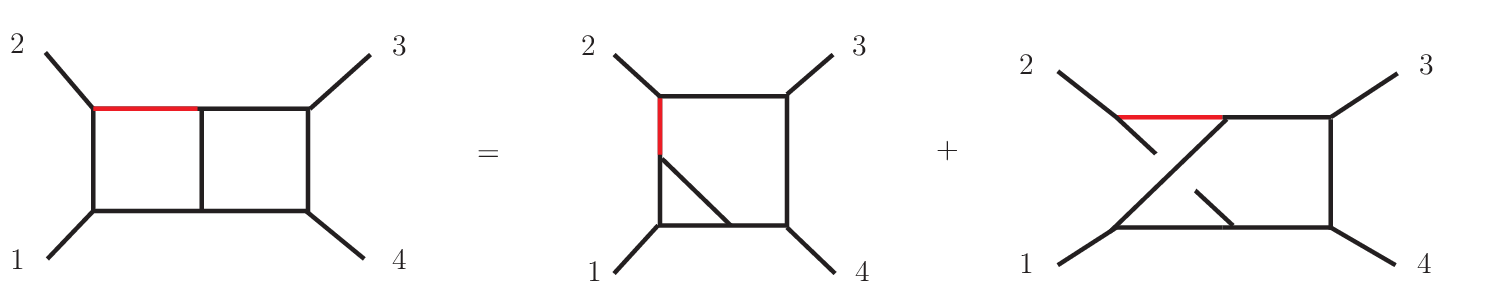} } 
	\caption{CK relation that should be excluded for deformations.} 
	\label{fig:2_loop_not_allowed_CK}
\end{figure}

In the off-shell CK relations discussed above, we have confined our focus to the topologies in Figure~\ref{fig:2_loop_deformed_topology}, ensuring that these relations do not extend to topologies that do not contribute to cut-(a).
For instance, the dual Jacobi relation shown in Figure~\ref{fig:2_loop_not_allowed_CK} should not be included because it involves topologies (the two on the right-hand side) that are absent in Figure~\ref{fig:2_loop_deformed_topology}.
In practice, one can achieve this by avoiding applying CK operation to the propagators that are severed by cut-(a), as this would lead to topologies outside the set requiring deformation.\footnote{Naturally, if $\Delta_i$ were allowed to propagate through all the topologies via dual Jacobi relations, we would revert to an integrand with global CK relations which makes no difference from the $n_i$ numerators.}

On the other hand, we require that the numerator $N_i$ upholds the global CK duality across a spanning set of cuts, ensuring the applicability of the double copy. Our construction for $\Delta_i$ is designed to maintain global CK duality specifically under cut-(a). Consequently, the relations we have excluded (such as Figure~\ref{fig:2_loop_not_allowed_CK}) imply that each $\Delta_i$ must vanish separately when subjected to cut-(b) and cut-(c). 
These additional constraints will be duly considered in our construction, as we demonstrate in the following sections.

\subsection{Explicit solution of deformation}\label{sec:deformsolution}

We now discuss the construction of the deformation $\Delta_i$.

\begin{figure}[t]
	\centerline{\includegraphics[height=3.7cm]{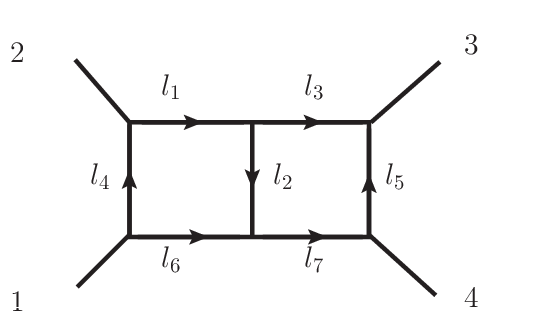} } 
	\caption{The double-box topology and its momentum labeling.} 
	\label{fig:double_box_diagram_two_labeled}
\end{figure}

To simplify the construction, it is convenient to divide the numerators into three parts according to the structure of polarization vectors.
For example, the numerators $n_i$ can be written as
\begin{equation}\label{eq:classification_of_terms}
	n_i = n_i^{[1]}+n_i^{[2]}+n_i^{[3]} \;.
\end{equation}
In $n_i^{[1]}$, each polarization vector is contracted with another polarization vector, for instance,
\begin{equation}
(\varepsilon_1 \cdot \varepsilon_2)(\varepsilon_3 \cdot \varepsilon_4)(p_1 \cdot p_2)(p_1 \cdot l_1)(p_1 \cdot l_2).
\end{equation}
Terms in  $n_i^{[2]}$ have two polarization vectors contracted with each other and the other two polarization vectors are contracted with momenta, such as 
\begin{equation}
(\varepsilon_1 \cdot \varepsilon_2)(\varepsilon_3 \cdot p_4)(\varepsilon_4 \cdot p_3)(p_1 \cdot l_1)(p_1 \cdot l_2).
\end{equation}
Finally, polarization vectors in $n_i^{[3]}$ are all contracted with momentas, and an example is
\begin{equation} 
(\varepsilon_1 \cdot p_2)(\varepsilon_2 \cdot p_3)(\varepsilon_3 \cdot p_4)(\varepsilon_4 \cdot p_3)(p_1 \cdot l_2).
\end{equation}
These three parts are independent with each other: they will not change the type during CK operations and they also satisfy unitarity cuts separately. Thus we can treat them independently. 
Following the same standard, the deformation $\Delta_i$ can also be divided into three parts: 
\begin{equation}\label{eq:classification_of_deformations}
 \Delta_i = \Delta_i^{[1]}+\Delta_i^{[2]}+\Delta_i^{[3]}.
\end{equation}
Below discuss these three parts one by one. 
Since the double-box topology, shown in Figure~\ref{fig:double_box_diagram_two_labeled}, is the master topology for the deformation, we only need to make an ansatz for $\Delta_1$.

\paragraph{Deformation $\Delta_1^{[1]}$.}

We consider $n_i^{[1]}$ and $\Delta_i^{[1]}$ first. Monominals in $n_i^{[1]}$ can be further devided into terms that are proportional to $(\varepsilon_1 \cdot \varepsilon_2)(\varepsilon_3 \cdot \varepsilon_4)$, $(\varepsilon_1 \cdot \varepsilon_3)(\varepsilon_2 \cdot \varepsilon_4)$ and $(\varepsilon_1 \cdot \varepsilon_4)(\varepsilon_2 \cdot \varepsilon_3)$, which should match the corresponding terms in the tree products respectively. 
We have calculated them separately, and we find that there is no difficulty for terms proportional to $(\varepsilon_1 \cdot \varepsilon_3)(\varepsilon_2 \cdot \varepsilon_4)$ and $(\varepsilon_1 \cdot \varepsilon_4)(\varepsilon_2 \cdot \varepsilon_3)$ in $n_i^{[1]}$ to match corresponding terms in tree products. The only inconsistency comes from terms propotional to $(\varepsilon_1 \cdot \varepsilon_2)(\varepsilon_3 \cdot \varepsilon_4)$, thus we can require all the terms in $\Delta_i^{[1]}$ to be propotional to $(\varepsilon_1 \cdot \varepsilon_2)(\varepsilon_3 \cdot \varepsilon_4)$. 

As mentioned in the end of Section~\ref{sec:generalidea}, the deformation should vanish under cut-(b) and cut-(c). A simple way to achieve this is to ask $\Delta_i^{[1]}$ to be proportional to some propagators which are cut by cut-(b) and cut-(c). For instance, we can acquire the deformation of double box topology $\Delta_1^{[1]}$ to be proportional to $(l_2)^2$ since either cut-(b) or cut-(c) will put it on shell and vanish.

Following the above discussion, we now can make an ansatz for $\Delta_1^{[1]}$:
\begin{equation}\label{ansatz_for_n_(1)}
	\Delta_1^{[1]} = (\varepsilon_1 \cdot \varepsilon_2)(\varepsilon_3 \cdot \varepsilon_4) (\sum_{k} c_{k}^{[1]} M^{[1]}_{k}) \, l_2^{2} \,,
\end{equation}
where $M^{[1]}_{k}$ are monominals formed by product of such basis:
\begin{equation}\label{n_(1)_basis}
	\{p_1 \cdot p_2, \; p_2 \cdot p_3, \; p_1 \cdot l_1, \; p_2 \cdot l_1, \; p_3 \cdot l_1, \; p_1 \cdot l_2, \; p_2 \cdot l_2, \; p_3 \cdot l_2, \; l_1 \cdot l_2, \; l_1^2, \; l_2^2\} 
\end{equation}
A simple dimension analysis shows that $M^{[1]}_{k}$ has mass dimension four. In this ansatz, we have 66 parameters in total. We can further constrain it by imposing symmetry conditions as in \eqref{eqs: double_box_symmetry}, and this will reduce parameters to 29. 

With the ansatz of $\Delta_1^{[1]}$, we can obtain other $\Delta_i^{[1]}$ by relations in \eqref{eq:2_loop_deformations_CK}, which satsify the symmetry properties automatically.
In addition, other $\Delta_i^{[1]}$ also vanish under cut-(b) and cut-(c) independently.
Finally, we match the expression with tree products of cut-(a). Indeed we find there are solutions,
and the 29 parameters in $\Delta_1^{[1]}$ will reduce to 28.

Interestingly, in the solution space, we find a special simple solution of $\Delta_1^{[1]}$ that can be given by a single term:
\begin{equation}\label{eq:deformed_numerators_n_(1)}
	\Delta_1^{[1]} = (d-2)^2(\varepsilon_1 \cdot \varepsilon_2)(\varepsilon_3 \cdot \varepsilon_4)l_4^2 \, l_2^2 \, l_5^2 \,,
\end{equation}
where $l_2$, $l_4$ and $l_5$ are labeled in Figure~\ref{fig:double_box_diagram_two_labeled}. 

We point out that in matching the cut conditions, one uses the full numerators $N_i$ given by \eqref{eq:deformed_numerators}, and thus the solution space of $n_i$ also receives further constraints from cut-(a) at the same time. Therefore, one has an updated solution of $n_i$ associated to the given solution of $\Delta_i$. In addition, in the above choice of the ansatz of $\Delta_i$, some degrees of freedom overlap with those in the solution space of $n_i$. This issue can be resolved by using a refined ansatz of $\Delta_i$ that is `orthogonal' to the solution space of $n_i$. We will discuss this in detail in Section~\ref{sec:solutionspaceofdeform}.

\paragraph{Deformation $\Delta_1^{[2]}$.}
As what we did for $n_{i}^{[1]}$, we carefully studied the origin of inconsistency in $n_{i}^{[2]}$. We find that the terms which can not match the tree product are either proportional to $\varepsilon_1 \cdot \varepsilon_2$ or $\varepsilon_3 \cdot \varepsilon_4$. So we make an ansatz for $\Delta_1^{[2]}$ as
\begin{equation}\label{eq:deformed_numerators_n_(2)}
	\Delta_1^{[2]} =\Big[ (\varepsilon_1 \cdot \varepsilon_2)(\sum_{a} c_{a}^{[2]} M^{[2]}_{1,a})+(\varepsilon_3 \cdot \varepsilon_4)(\sum_{b} c_{b}^{[2]} M^{[2]}_{2,b}) \Big] l_2^{2} \,,
\end{equation}
where the corresponding basis for the monomials $M^{[2]}_{1,a}$ and $M^{[2]}_{2,b}$ are
\begin{equation}\label{eq:basis_for_deformed_n_(2)}
	\{\varepsilon_k \cdot p_i,\; \varepsilon_k \cdot l_{\alpha},\;
	p_i \cdot p_j, \; p_i \cdot l_{\alpha}, \;  l_{\alpha} \cdot l_{\beta}\},
\end{equation}
where $k= 1,2$ for $M^{[2]}_{2,b}$ and $k=3,4$ for $M^{[2]}_{1,a}$. Note that the ansatz is proportional to $(l_2)^2$ so that it vanishes under cut-(b) and cut-(c). In this ansatz $\Delta_1^{[2]}$ contains 352 parameters and symmetry conditions will reduce it to 102. Other $\Delta_i^{[2]}$ will be determined by the relations in \eqref{eq:2_loop_deformations_CK}, and they also satisfy the symmetry properties and vanish under cut-(b) and cut-(c).

Finally, we impose the cut-(a) constraint for the deformed integrand based on $N_{i}^{[2]}$ and now the solution exists. We find that the 102 parameters in $\Delta_1^{[2]}$ reduce to 100.
Similar to $\Delta_1^{[1]}$, we find a very simple special solution for $\Delta_1^{[2]}$ within the solution space:
\begin{equation}\label{eq:Delta(2)_simple_form}
	\Delta_1^{[2]} = -4(d-2)^2\Big[(\varepsilon_1 \cdot \varepsilon_2)(\varepsilon_3 \cdot l_{5})(\varepsilon_4 \cdot l_{5})l_4^2 + (\varepsilon_3 \cdot \varepsilon_4)(\varepsilon_1 \cdot l_{4})(\varepsilon_2 \cdot l_{4})l_5^2 \Big]l_2^2.
\end{equation}

\paragraph{Deformation $\Delta_1^{[3]}$.}
Finally, we determine the $\Delta_1^{[3]}$. Inspecting the structure of $\Delta_1^{[1]}$ and $\Delta_1^{[2]}$, one notes that $\varepsilon_1$ and $\varepsilon_2$ only contract with $l_{4}$, while $\varepsilon_3$ and $\varepsilon_4$ only contract with $l_{5}$. And a naive guess for the minimal ansatz of $\Delta_1^{[3]}$ could be
\begin{equation}\label{eq:deformed_numerators_Delta(3)_naive}
	\Delta_1^{[3]} = c_1^{[3]}(\varepsilon_1 \cdot l_{4})(\varepsilon_2 \cdot l_{4})(\varepsilon_3 \cdot l_{5})(\varepsilon_4\cdot l_{5})l_2^2.
\end{equation}
Unfortunately, we find deformation $\Delta_1^{[3]}$ in this simple form can not pass all the unitarity cuts, which means we need to enlarge the ansatz and provide a more general form. 

We propose $\Delta_1^{[3]}$ as: 
\begin{equation}
	\label{ansatz}
	\Delta_1^{[3]}  =  (\sum_{k} c_{k}^{[3]} M^{[3]}_{k})l_2^2\;, \,
\end{equation}
where $M^{[3]}_k$ is formed by the product of following basis:
\begin{equation}
	\{\varepsilon_i \cdot p_j,\;\varepsilon_i \cdot l_{\alpha} \}
\end{equation}
with $i,j = 1,2,3,4$ , $\alpha = 1,2$. As before, the ansatz is proportional to $(l_2)^2$ so that it vanishes under cut-(b) and cut-(c). 
In this ansatz, there are 256 parameters and after imposing the symmetry constraints, 76 parameters will remain.

Again, we find the deformed integrand $N_i^{[3]}$ can pass all the unitarity cuts, and the 76 parameters in $\Delta_1^{[3]}$ will reduce to 65. 

Within the solution space, we present a relatively simple specific solution for $\Delta_1^{[3]}$:
\begin{equation}\label{eq:deformed_numerators_n_(3)}
	\Delta_1^{[3]}=4(d-2)\Big[ 4(d-2)S_1^{[3]} -10 S_2^{[3]} -10 S_3^{[3]} - 20 S_4^{[3]} + 47 S_5^{[3]} -32 S_6^{[3]} \Big] l_2^2,
\end{equation}
where
\begin{equation}\label{Delta(3)_symmetry_basis}
	\begin{aligned}
		S_1^{[3]}&=(\varepsilon_1 \cdot l_4)(\varepsilon_2 \cdot l_4)(\varepsilon_3 \cdot l_5)(\varepsilon_4 \cdot l_5)\\
		S_2^{[3]}&=(\varepsilon_1 \cdot l_5)(\varepsilon_2 \cdot l_5)(\varepsilon_3 \cdot l_4)(\varepsilon_4 \cdot l_4)\\
		S_3^{[3]}&=(\varepsilon_1 \cdot l_1)(\varepsilon_2 \cdot l_6)(\varepsilon_3 \cdot l_7)(\varepsilon_4 \cdot l_3)\\
		S_4^{[3]}&=(\varepsilon_1 \cdot l_5)(\varepsilon_2 \cdot l_6)(\varepsilon_3 \cdot l_4)(\varepsilon_4 \cdot l_3)+(\varepsilon_1 \cdot l_1)(\varepsilon_2 \cdot l_5)(\varepsilon_3 \cdot l_7)(\varepsilon_4 \cdot l_4)\\
		S_5^{[3]}&=(\varepsilon_1 \cdot l_7)(\varepsilon_2 \cdot l_3)(\varepsilon_3 \cdot l_7)(\varepsilon_4 \cdot l_3)+(\varepsilon_1 \cdot l_1)(\varepsilon_2 \cdot l_6)(\varepsilon_3 \cdot l_1)(\varepsilon_4 \cdot l_6)\\
		S_6^{[3]}&=(\varepsilon_1 \cdot p_3)(\varepsilon_2 \cdot p_4)(\varepsilon_3 \cdot l_3)(\varepsilon_4 \cdot l_7)+(\varepsilon_1 \cdot l_6)(\varepsilon_2 \cdot l_1)(\varepsilon_3 \cdot p_1)(\varepsilon_4 \cdot p_2) \,,
	\end{aligned}
\end{equation}
where the loop momenta $l_i$ are labeled in Figure~\ref{fig:double_box_diagram_two_labeled}.     

Combining the solutions of the deformation together with $n_i$, we obtain the complete integrand with $N_i$ which passes all unitarity cuts. Since they satisfy the CK-dual relations on the cuts, they can be directly used to construct the gravitational amplitude via double copy.

\begin{figure}[t]
	\centerline{\includegraphics[height=2.6cm]{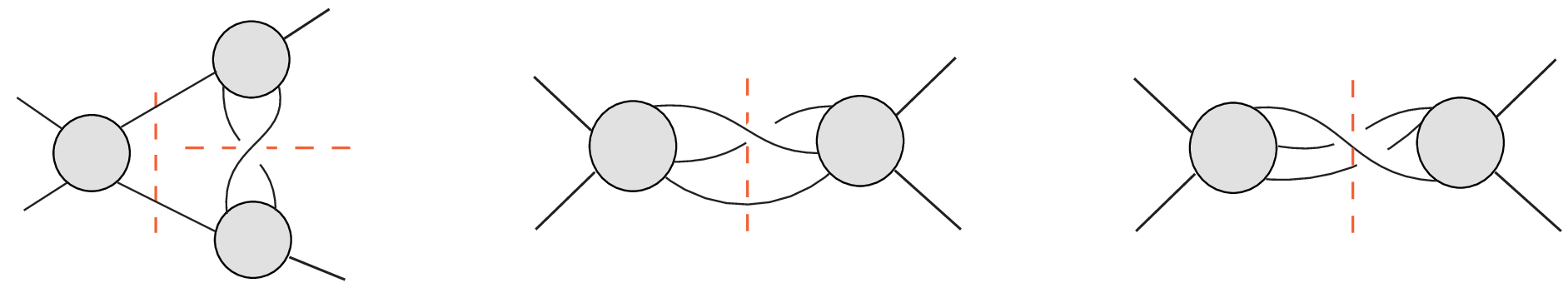} } 
	\caption{Non-planar cuts of the two-loop four-point amplitude. } 
	\label{2_loop_non_planar_cut}
\end{figure}

Let us mention a few further checks of the results. First, as mentioned before, in the above construction one can use only planar cuts. We have checked that the solutions obtained in this way automatically satisfy the full set of non-planar cuts as shown in Figure~\ref{2_loop_non_planar_cut}.
Second, we perform the projection of the integrand into a set of gauge invariant bases (see \emph{e.g.}~\cite{Boels:2018nrr}), followed by the IBP reduction into a set of independent master integrals \cite{Chetyrkin:1981qh, Tkachov:1981wb}, and we find all the free parameters cancel out. 
Finally, after integration, we have checked that the full-color IR divergences are consistent with the known Catani formula \cite{Catani:1998bh}. 
We provide the numerator solution and other factors of the integrand in the ancillary files.

 \subsection{Understanding the solution space of deformations}\label{sec:solutionspaceofdeform}
 
In the last subsection, we introduced the method of adding deformations to the global CK-dual numerators to make it pass all the unitarity cuts. 
By analyzing the structure of deformations, we can write down a general ansatz for the deformed numerators $\Delta_i$. 
After passing all unitarity cuts, we find there remain free parameters. 
In particular, the free parameters in $\Delta_i$ have redundancy that overlaps with the free parameters in $n_i$.

A natural question arises of whether we can find the space of $\Delta_i$ that is orthogonal to the $n_i$ solution space at the beginning. This will not only simplify the ansatz but also help us to understand the structure of deformation. 
In this subsection, we discuss this problem and show how to construct the orthogonal space for $\Delta_i$.

\paragraph{Deformation $\Delta_1^{[1]}$.}
Again we start with $\Delta_1^{[1]}$ to illustrate the strategy. We recall the general form of the ansatz given in \eqref{ansatz_for_n_(1)}. Since we require it to reflect the symmetry of the double box diagram, also to simplify the discussion below, we will expand the ansatz as a linear combination of a set of symmetric bases as\footnote{The choice of the symmetry bases is not unique.}
\begin{equation}\label{Delta1_symmetry}
	\Delta_1^{[1]}= (\varepsilon_1 \cdot \varepsilon_2)(\varepsilon_3 \cdot \varepsilon_4) l_2^{2} \sum_{i=1}^{29}c_{i}^{[1]}S_{i}^{[1]},
\end{equation}
where each $S_{i}^{[1]}$ satisfies the symmetry properties, for example, the first two $S_{i}^{[1]}$ are
\begin{equation}\label{eq:symbasis12}
	\begin{aligned}
		S_1^{[1]}&=l_4^2 \, l_5^2 \,, \\
		S_2^{[1]}&=(l_1 \cdot p_3)(l_2 \cdot p_1)-(l_2  \cdot p_4)(l_3 \cdot p_2)- (l_2 \cdot p_2)(l_6 \cdot p_4) + (l_2 \cdot p_3)(l_7 \cdot p_1) \,.
	\end{aligned}
\end{equation}
One can check that they satisfy the symmetry relations  as in \eqref{eqs: double_box_symmetry}.

Our goal is to reduce \eqref{Delta1_symmetry} so that it is orthogonal to the solution space of $n_1^{[1]}$.
For this, we need to analyze the solution space of $n_1^{[1]}$ carefully. 
We recall that, in obtaining the solution space of $n_1$, we have imposed the global CK-duality relations, the graph symmetries, as well as the cut-(b) and cut-(c).

We note that all the symmetry basis $S_{i}^{[1]}$ in \eqref{Delta1_symmetry} also appear in the original ansatz of $n_1^{[1]}$. We extract this part in the solution space of $n_1$ as
\begin{equation}\label{eq:n1delta1}
	n_1^{\Delta,[1]}=  (\varepsilon_1 \cdot \varepsilon_2)(\varepsilon_3 \cdot \varepsilon_4) l_2^{2} \sum_{i=1}^{29}c_{i}^{\prime [1]}S_{i}^{[1]} \,,
\end{equation}
where the parameters $c_{i}^{\prime [1]}$ in \eqref{eq:n1delta1} are not all independent but satisfy certain linear relations.

An important point to note is that it is these constraints for $c_{i}^{\prime [1]}$ that make the numerator solutions $n_i$ incompatible with cut-(a). 
Therefore, the deformation that we introduce should play the role of relaxing these constraints, 
and the ansatz that exactly  relaxes these constraints is what we call \emph{orthogonal} to the solution space of $n_1$.

To find the minimal orthogonal space, there is one further redundancy that we can remove.
Namely, the basis $S_{i}^{[1]}$ can be equivlent under cut-(a). Concretely, we define the contribution to cut-(a) associated to certain $S_{i}^{[1]}$ as
\begin{equation}
	\mathcal{I}_{\textrm{cut(a)}}(S_{i}^{[1]})= \Bigg( l_2^2 S_{i}^{[1]} 
	\begin{gathered}
		\includegraphics[height=2cm]{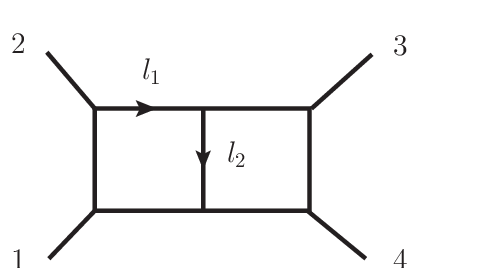}
	\end{gathered}
	\hskip-.5cm+ \  \textrm{CK-induced contribution} \Bigg)\Bigg|_{\textrm{cut-(a)}}.
\end{equation}
Here ``CK-induced contribution" represents all the terms generated from other topologies through the set of dual Jacobi relations in \eqref{eq:2_loop_deformations_CK} by taking $\Delta_1 = l_2^2 S_{i}^{[1]}$. If $\mathcal{I}_{\textrm{cut(a)}}(S_{i}^{[1]})$ is equal to 0, we simply drop the corresponding $S_{i}^{[1]}$ in $n_1^{\Delta,[1]}$. 
And if $\sum_i \mathcal{I}_{\textrm{cut(a)}}(S_{i}^{[1]})=0$, we  can solve for one of $S_{i}^{[1]}$ in terms of other symmetry basis  in $n_1^{\Delta,[1]}$. 
After eliminating these redundancies, we find that $n_1^{\Delta,[1]}$ becomes:
\begin{equation}
	\widetilde{n}_1^{\Delta,[1]}=l_2^{2} \sum_{i=1}^{10} {\tilde c}_{i}^{[1]}S_{i}^{[1]} \,.
\end{equation}
Note that $\widetilde{n}_1^{\Delta,[1]}$ is equivalent to $n_1^{\Delta,[1]}$ under cut-(a).

We also obtain the coefficients ${\tilde c}_{i}^{[1]}$ in the solution space of $n_1$, and the key point to analyze their linear relations.
We find there exists one and only one linear relation that is among the two bases in \eqref{eq:symbasis12}:
\begin{equation}\label{eq:c1eqc2}
	{\tilde c}_{1}^{[1]}=-{\tilde c}_{2}^{[1]},
\end{equation}
and other ${\tilde c}_{i}^{[1]}$ are independent with each other.  
The reason that the global CK integrand $n_i^{[1]}$ can not pass cut-(a) is due to that this linear relation is incompatible with cut-(a).

Therefore, the deformation $\Delta_1^{[1]}$ should play a role to relax this condition.
It is sufficient to make an ansatz for $\Delta_1^{[1]}$ with only one term:
\begin{equation}
	\Delta_1^{[1]}= (\varepsilon_1 \cdot \varepsilon_2)(\varepsilon_3 \cdot \varepsilon_4)(c_1^{[1]}S_1^{[1]}) \,  l_2^{2} \,.
\end{equation}
Note that $c_1^{[1]}$ is a free parameter, and by adding this deformation, one relaxes the constraint \eqref{eq:c1eqc2} in the full numerator $N_i^{[1]} = n_i^{[1]} + \Delta_i^{[1]}$.
Now one finds that the ansatz indeed has a solution under cut-(a) which gives
\begin{equation}
c_1^{[1]} = (d-2)^2 \,.
\end{equation}
Note that this term is exactly what we find in \eqref{eq:deformed_numerators_n_(1)}. 

Alternatively, we can also choose the $S_2^{[1]}$ as the deformation $\Delta_1^{[1]}$ since it will play the same role of relaxing the constraint \eqref{eq:c1eqc2}. 
We can think that $S_1^{[1]}$ or $S_2^{[1]}$ forms the complete orthogonal space for $\Delta_1^{[1]}$.
In this way, we obtain the minimal ansatz space for $\Delta_1^{[1]}$.

\paragraph{Deformation $\Delta_1^{[2]}$.}
As in the previous case, we extract the corresponding 102 bases of $\Delta_1^{[2]}$ in the solution space of $n_{1}^{[2]}$:  
\begin{equation}\label{n1_Delta_2}
	n_1^{\Delta,[2]}=  l_2^{2} \sum_{i=1}^{102}c_{i}^{\prime [2]}S_{i}^{[2]} \,,
\end{equation}
where we have organized it on a basis that respects the symmetry property.
After removing the redudancey from $\mathcal{I}_{\textrm{cut(a)}}(S_{i}^{[2]})$, we get the numerator with 32 symmetry bases:
\begin{equation}
	\widetilde{n}_1^{\Delta,[2]}=l_2^{2} \sum_{i=1}^{32} {\tilde c}_{i}^{[2]}S_{i}^{[2]}.
\end{equation}
The coefficients ${\tilde c}_{i}^{[2]}$ are determined by the solution space of $n_1$. As before, the key point to analyze their linear relations.
We find two linear relations for ${\tilde c}_{i}^{[2]}$:
\begin{equation}\label{eq:crelationDelta2}
	\begin{aligned}
		{\tilde c}_{1}^{[2]}&=-\frac{1}{2}({\tilde c}_{3}^{[2]}+{\tilde c}_{4}^{[2]}+{\tilde c}_{5}^{[2]}+{\tilde c}_{6}^{[2]}) \,,\\
		{\tilde c}_{2}^{[2]}&=\frac{1}{2}({\tilde c}_{8}^{[2]}-{\tilde c}_{7}^{[2]}) \,.\\
	\end{aligned}
\end{equation}
To relax these two relations, it is enough to introduce the deformation $\Delta_1^{[2]}$ as:
\begin{equation}
	\Delta_1^{[2]}=l_2^2(c_1^{[2]}S_1^{[2]}+c_2^{[2]}S_2^{[2]}),
\end{equation}
and we give $S_1^{[2]}$ and $S_2^{[2]}$ explicitly as
\begin{equation}
	\begin{aligned}
		S_1^{[2]}&=(\varepsilon_1 \cdot \varepsilon_2)(\varepsilon_3 \cdot l_5)(\varepsilon_4 \cdot l_5)l_4^2 + (\varepsilon_3 \cdot \varepsilon_4)(\varepsilon_1 \cdot l_4)(\varepsilon_2 \cdot l_4)l_5^2 \,, \\
		S_2^{[2]}&=(\varepsilon_1 \cdot \varepsilon_2)((\varepsilon_3 \cdot l_6)(\varepsilon_4 \cdot p_1) + (\varepsilon_3 \cdot p_2)(\varepsilon_4 \cdot l_1))l_5^2 \\
		& \qquad -(\varepsilon_3 \cdot \varepsilon_4)((\varepsilon_1 \cdot l_3)(\varepsilon_2 \cdot p_3) + (\varepsilon_1 \cdot p_4)(\varepsilon_2 \cdot l_7))l_4^2 \,.
	\end{aligned}
\end{equation}
By matching the cut-(a) we can solve for $c_1^{[2]}$, $c_2^{[2]}$:
\begin{equation}
c_1^{[2]} = -4(d-2)^2 \,, \qquad c_2^{[2]} = 0\,.
\end{equation}
This solution is equivalent to the form in \eqref{eq:Delta(2)_simple_form}.

One is free to make a different ansatz by choosing different linear combinations of two $S_i^{[2]}$ from the relation \eqref{eq:crelationDelta2}, for instance, $\Delta_1^{[2]}=l_2^2(c_3^{[2]}S_3^{[2]}+c_8^{[2]}S_8^{[2]})$ will be a perfect ansatz for the deformation as well. The explicit expressions of the symmetry basis $S_{i}^{[2]}, i=3,\ldots, 8,$ are provided in the ancillary file.

\paragraph{Deformation $\Delta_1^{[3]}$.}
Finally, for $\Delta_1^{[3]}$ we also extract the terms proportional to $l_2^2$ and repeat the above operation. 
After removing the redundancy under cut-(a), 
39 symmetry bases remain and one has
\begin{equation}
	\widetilde{n}_1^{\Delta,[3]}=l_2^2\sum_{i=1}^{39}{\tilde c}_{i}^{[3]}S_{i}^{[3]} \,.
\end{equation}
Again, ${\tilde c}_{i}^{[3]}$ are determined by the solution space of $n_1$, and one finds that there are 11 linear relations for ${\tilde c}_{i}^{[3]}$. 
To relax these 11 relations, one can pick one basis from each relation and form a linear ansatz of $\Delta_1^{[3]}$ as
\begin{equation}
	\Delta_1^{[3]}=l_2^2\sum_{i=1}^{11}c_{i}^{[3]}S_{i}^{[3]},
\end{equation}
and after matching cut-(a), all the coefficients $c_{i}^{[3]}$ can be uniquely fixed. 
Using the same symmetry basis including those in \eqref{Delta(3)_symmetry_basis}, we reproduce the previously obtained solution \eqref{eq:deformed_numerators_n_(3)}.

We provide the details of the 11 linear relations of  ${\tilde c}_{i}^{[3]}$ and the involved symmetry basis $S_{i}^{[3]}$ in the ancillary file.
Using them, one can construct different ansatz for $\Delta_1^{[3]}$.

\section{Summary}
\label{sec:discussion}

 In this paper, we revisit the color-kinematics (CK) duality of the two-loop four-gluon amplitudes in pure non-supersymmetric Yang-Mills theory.
 We observe that the local-form ansatz fails to simultaneously satisfy global CK duality and unitarity cuts. 
 However, we identify a simple deformation that rectifies this issue.
This approach allows us to leverage the off-shell CK-dual relations to the fullest extent while keeping the ansatz compact during the construction process. 
 
The final numerators can be given as
\begin{equation}
	N_i = \left\{ \begin{matrix} & n_i +\Delta_i, & \qquad\qquad i \in \{\textrm{cut(a)-related topologies}\} \\ & n_i , & \textrm{other topologies.} \end{matrix} \right.
\end{equation} 
Here $n_i$ satisfy the complete set of dual Jacobi relations for all topologies in \eqref{eq:dualjacobirelationforni}, and $\Delta_i$ represent the deformation that fulfills a subset of dual Jacobi relations  \eqref{eq:2_loop_deformations_CK} within the cut(a)-related topologies. 
All these Jacobi relations are off-shell, meaning they are not subject to any cut constraints.
To derive the $n_i$ part of the numerators, only two master numerators $n_1$ and $n_2$ are required. 
For $\Delta_i$, only one master numerator $\Delta_1$ (for the subset relations) is needed. 
A particular solution of the deformation for $\Delta_1$ can be given in the following remarkably simple form:
\begin{align} \label{eq:deformed_1}
	 \Delta & \left[  \begin{gathered} {\includegraphics[height=1.9cm]{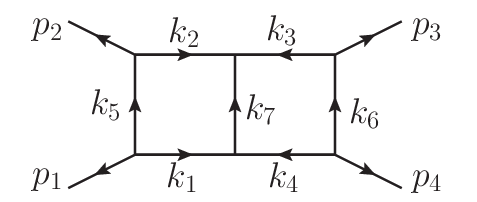}} \end{gathered} \right] \bigg/ k_7^2 = \\
	&+ (d-2)^2 \bigg\{ (\varepsilon_1 \cdot \varepsilon_2)(\varepsilon_3 \cdot \varepsilon_4) k_5^2 \, k_6^2 
	 +16 (\varepsilon_1 \cdot k_5)(\varepsilon_2 \cdot k_5)(\varepsilon_3 \cdot k_6)(\varepsilon_4 \cdot k_6) \nonumber\\
	& \qquad\qquad\quad -4 
	\Big[(\varepsilon_1 \cdot \varepsilon_2)(\varepsilon_3 \cdot k_6)(\varepsilon_4 \cdot k_6) k_5^2 + (\varepsilon_3 \cdot \varepsilon_4)(\varepsilon_1 \cdot k_5)(\varepsilon_2 \cdot k_5) k_6^2 \Big] \bigg\} \nonumber\\
	&  + (d-2) 4 \bigg\{ -10 \Big[ (\varepsilon_1 \cdot k_6)(\varepsilon_2 \cdot k_6)(\varepsilon_3 \cdot k_5)(\varepsilon_4 \cdot k_5) 
	+ (\varepsilon_1 \cdot k_2)(\varepsilon_2 \cdot k_1)(\varepsilon_3 \cdot k_4)(\varepsilon_4 \cdot k_3) \Big] \nonumber\\
	& \qquad\qquad\qquad  + 20 
	\Big[(\varepsilon_1 \cdot k_6)(\varepsilon_2 \cdot k_1)(\varepsilon_3 \cdot k_5)(\varepsilon_4 \cdot k_3)+(\varepsilon_1 \cdot k_2)(\varepsilon_2 \cdot k_6)(\varepsilon_3 \cdot k_4)(\varepsilon_4 \cdot k_5) \Big] \nonumber\\
	& \qquad\qquad\qquad +  32 \Big[ (\varepsilon_1 \cdot k_5)(\varepsilon_2 \cdot k_5)(\varepsilon_3 \cdot p_1)(\varepsilon_4 \cdot p_2)+ (\varepsilon_1 \cdot p_3)(\varepsilon_2 \cdot p_4)(\varepsilon_3 \cdot k_6)(\varepsilon_4 \cdot k_6)\Big]  \nonumber\\
	& \qquad\qquad\qquad + 47 \Big[ (\varepsilon_1 \cdot k_4)(\varepsilon_2 \cdot k_3)(\varepsilon_3 \cdot k_4)(\varepsilon_4 \cdot k_3)+(\varepsilon_1 \cdot k_2)(\varepsilon_2 \cdot k_1)(\varepsilon_3 \cdot k_2)(\varepsilon_4 \cdot k_1) \Big] \bigg\} , \nonumber
\end{align}
which is a sum of the terms in \eqref{eq:deformed_numerators_n_(1)}, \eqref{eq:Delta(2)_simple_form} and \eqref{eq:deformed_numerators_n_(3)} (note that we have relabeled the loop momenta to make the symmetry property more manifest).
The simplicity of the deformation suggests that the violation of global off-shell CK duality is, in fact, quite limited.
To fully appreciate the simplicity of the above result, the reader should compare this formula with the much more complex expression of the $n_1$ component of the numerator, which is provided in the ancillary files.

We also discuss in detail how to construct the minimal ansatz of $\Delta_i$ in the ``orthogonal space" relative to the solution space of $n_i$.  
This can further narrow the parameter space of the deformation significantly.
The solution we obtain, while partially breaking the off-shell dual Jacobi relations, does satisfy the global CK-dual relations on all cuts, therefore, they can be directly used to construct the corresponding gravitational amplitude through double copy as in \cite{Bern:2015ooa}.

Our strategy provides a new efficient method to construct loop integrands via CK duality by maximizing using the off-shell CK-duality.
Notably, the deformation itself also satisfies a subset of dual Jacobi relations. 
Consequently, our approach requires far fewer parameters compared to previous constructions. 
For comparison, in \cite{Bern:2015ooa} where CK duality is imposed only on the cut integrand, the ansatz has to be made for the numerators of all trivalent topologies, and there are 6322 parameters remaining in the final solution. On the other hand, our construction focuses solely on the numerators of master topologies, and the complexity is similar to the usual effective construction based on global CK duality.

It would be interesting to apply our strategy to other high loop amplitudes or form factors that have been challenging to construct using CK duality. 
A pertinent question is whether the simplicity of the deformation is a general characteristic. Furthermore, could there be an underlying mathematical or physical principle governing the deformation as indicated by \eqref{eq:deformed_1}? A better understanding of these questions may emerge with more data available. We leave these intriguing problems for future investigation.

\acknowledgments

We would like to thank Bo Feng and Qingjun Jin for the discussion. 
This work is supported in part by the National Natural Science Foundation of China (Grants No.~12175291, 11935013, 12047503, 12247103) 
and by the CAS under Grants No.~YSBR-101.
We also thank the support of the HPC Cluster of ITP-CAS.

\providecommand{\href}[2]{#2}\begingroup\raggedright\endgroup

\end{document}